\documentclass{kapproc} 

\usepackage{amsmath, amssymb}

\usepackage{procps}

\usepackage[dvips]{graphicx}

\upperandlowercase

\kluwerbib 

\begin{document}

\articletitle[RELATIVISTIC SHOCK WAVES WITH LOSSES IN GRB SOURCES]{ON DYNAMICS OF RELATIVISTIC SHOCK WAVES WITH LOSSES IN GAMMA-RAY BURST SOURCES}


\author{E.V. Derishev, Vl.V. Kocharovsky, K.A. Martiyanov}
\affil{Institute of Applied Physics\\ 46 Ulyanov st., 603950
Nizhny Novgorod, Russia} \email{mca1@appl.sci-nnov.ru}

\begin{abstract}
Generalization of the self-similar solution for ultrarelativistic
shock waves \cite{BM} is obtained in presence of losses localized
on the shock front or distributed in the downstream medium. It is
shown that there are two qualitatively different regimes of shock
deceleration, corresponding to small and large losses. We present
the temperature, pressure and density distributions in the
downstream fluid as well as Lorentz factor as a function of
distance from the shock front.
\end{abstract}



\section*{Introduction}

The progenitors of gamma-ray bursts (GRBs) are believed to produce
highly relativistic shocks at the interface between the ejected
material and ambient medium (see, e.g., \cite{Mesz}; \cite{Piran}
for review). Non-thermal spectra and short duration of GRBs place
a firm lower limit to the bulk Lorentz factor of radiating plasma,
which must exceed a few hundred to avoid the compactness problem
(e.g., \cite{Baring}).

Consider a relativistic spherical blast wave expanding into a
uniform ambient medium with the Lorentz factor $\Gamma \sim 300$.
The average energy per baryon in the fluid comoving frame behind
the shock front is of the order of $\Gamma m_p c^2$ \cite{Taub},
where $m_p$ is proton mass, and the plasma in the downstream
presumably forms a non-thermal particle distribution extending up
to very high energies. Under these conditions, medium downstream
is subject to various loss processes. The non-thermal electrons
produce synchrotron radiation, which accounts for GRB afterglow
emission, and (at least partially) for the prompt emission. Apart
from the synchrotron radiation of charged particles there is
another mechanism of energy and momentum losses connected with
inelastic interactions of energetic protons with photons. These
reactions cause proton-neutron conversion as a result of charged
pion creation. It should be noticed that for typical interstellar
density the Coulomb collisions are inefficient and the charged
particles instead interact collectively through the magnetic
field. This allows to describe plasma motion using hydrodynamical
approach, though it can break for a small fraction of the most
energetic particles.

When a proton turns into a neutron or another neutral particle is
born, it does not interact with the magnetic field and hence the
energy spent for its creation is lost from the hydrodynamical
point of view. The synchrotron and inverse Compton emission, as
well as energetic photons, neutrinos and neutrons produced via
photopionic reactions, escape from downstream giving rise to
non-zero divergence of the energy-momentum tensor. The creation of
energetic neutrons is also a first step in the production of
highest-energy cosmic rays through the converter mechanism
(\cite{DKK}).

Ejection from the GRB progenitor of a mass $M_0$ with initial
Lorentz factor $\Gamma_0$ results in two shocks propagating
asunder from the contact discontinuity. The forward shock moves
into the external gas and has a much greater compression ratio at
its front than the other, reverse shock, which passes through the
ejected matter. As the shocked external gas has a temperature much
higher than that in the vicinity of the reverse shock, we neglect
the losses in the ejecta.

We discuss two models. In the first one we assume the energy
losses to be localized close to the shock front, whereas the
matter downstream the shock is considered lossless. In another
model the shock front is treated as non-dissipative and the losses
are distributed all over the shocked gas.

Following the recipe of Blandford and McKee (1976) we generalize
their well-known self-similar solutions for relativistic blast
waves for the case, where the energy and momentum of the
relativistic fluid is carried away by various species of neutral
particles.

\section{Self-similar solutions}

We start from the energy-momentum continuity equations, where in
the case of distributed losses a non-zero r.h.s. is
included:\\[-2\smallskipamount]
$$%
\frac{\partial T^{00}}{c\partial t}+\frac{1}{r^2}\frac{\partial
(r^2 T^{0r})}{\partial r}=-\varphi_0\,T^{00}, \ \
\frac{\partial T^{0r}}{c\partial t}+\frac{1}{r^2}\frac{\partial
(r^2 T^{rr})}{\partial r}-
            \frac{2p}{r}=-\varphi_1\,T^{0r},
$$%
$$%
T^{00}=w\gamma^2-p,\qquad T^{0r}=w\gamma^2\beta,\qquad
T^{rr}=w\gamma^2\beta^2+p,
$$%
where $T$ is the energy-momentum tensor, $\gamma$ the Lorentz
factor, $w{=}e{+}p$ the enthalpy density, $p$ the pressure, $e$
the energy density. All quantities are measured in the fluid
comoving frame. In the following analysis we use the
ultrarelativistic approximation of these equations, obtained by
expanding velocity up to the third contributing order in
$\gamma^{-2}$ and the equation of state up to the first order.
Because of the lack of space, here we consider only equal losses
for the energy and momentum ($\varphi_0 {=} \varphi_1 {=}
\varphi$).

\begin{figure}[!t]
\centering
\abovecaptionskip=-10pt
\parbox[t]{.54\textwidth}{ \includegraphics[width=.41\textwidth]{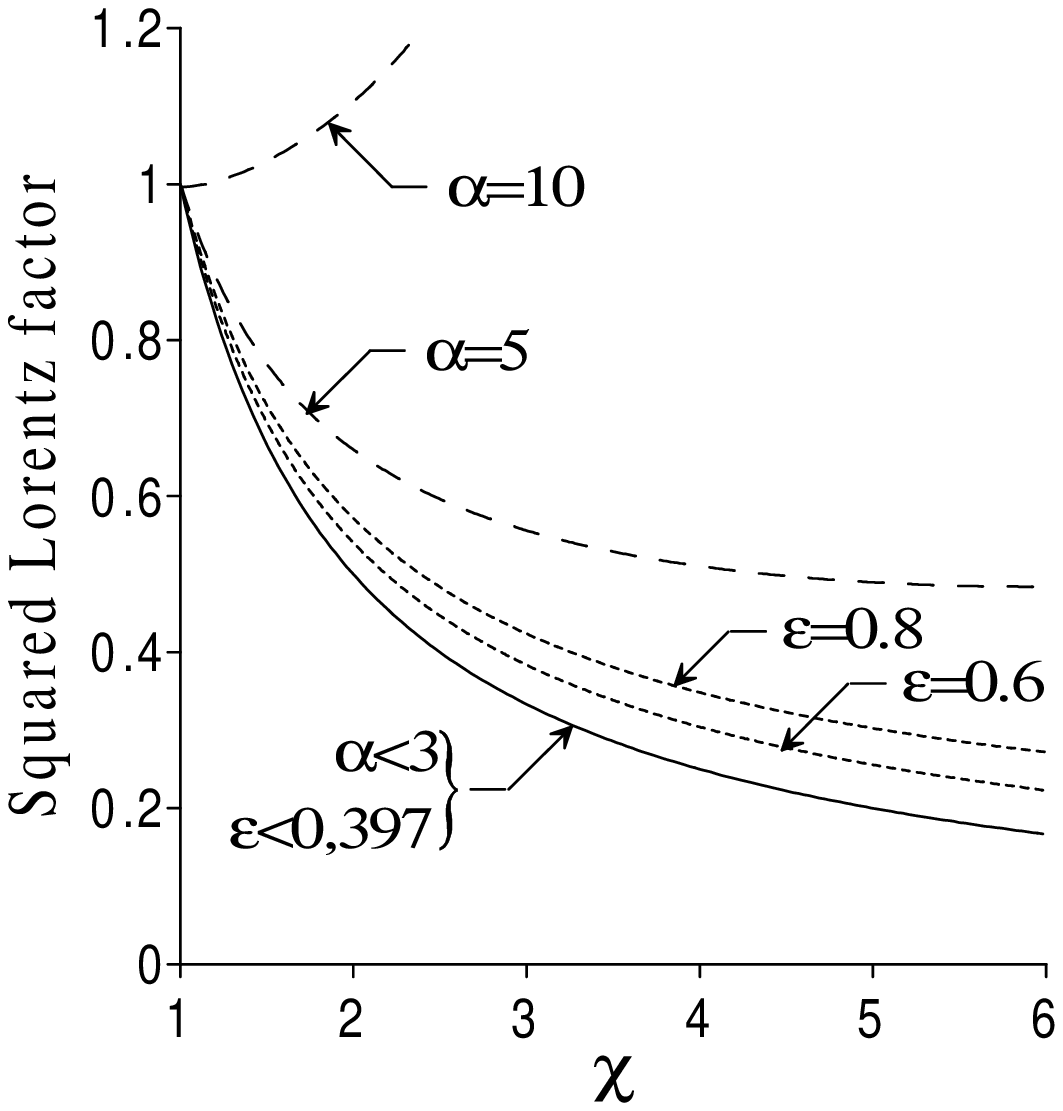}
}%
\hfill
\parbox[t]{.45\textwidth}{ \includegraphics[width=.41\textwidth]{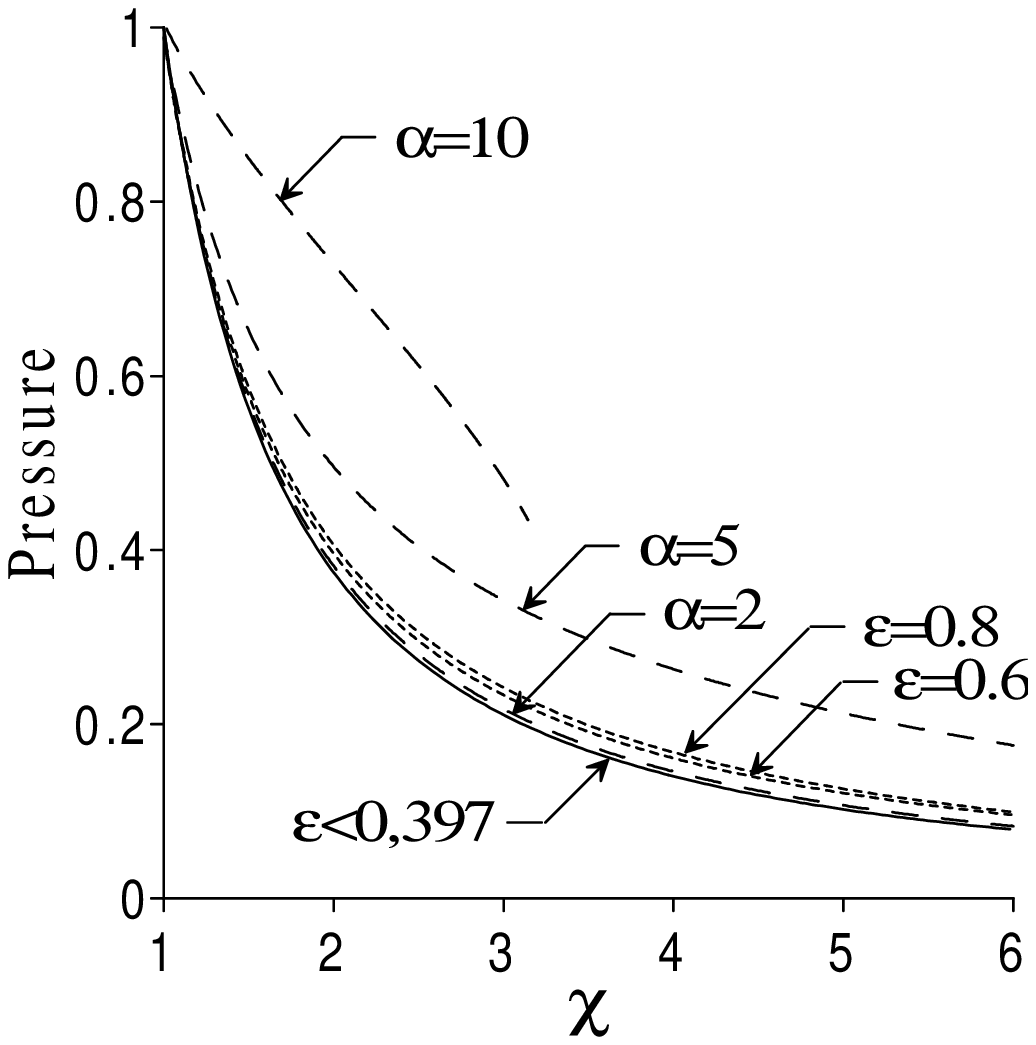}
}\\
\parbox[t]{.54\textwidth}{ \includegraphics[width=.41\textwidth]{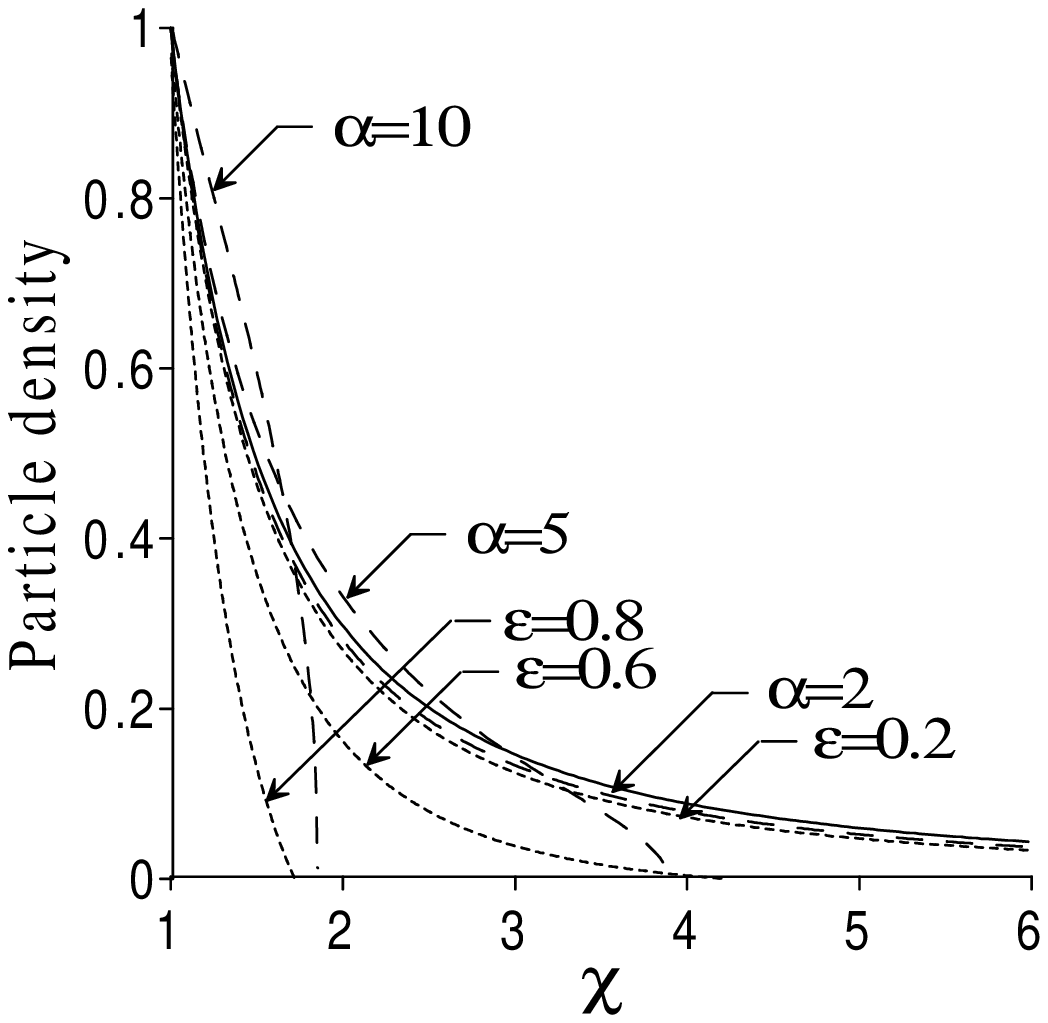}
}%
\hfill
\parbox[t]{.45\textwidth}{ \includegraphics[width=.41\textwidth]{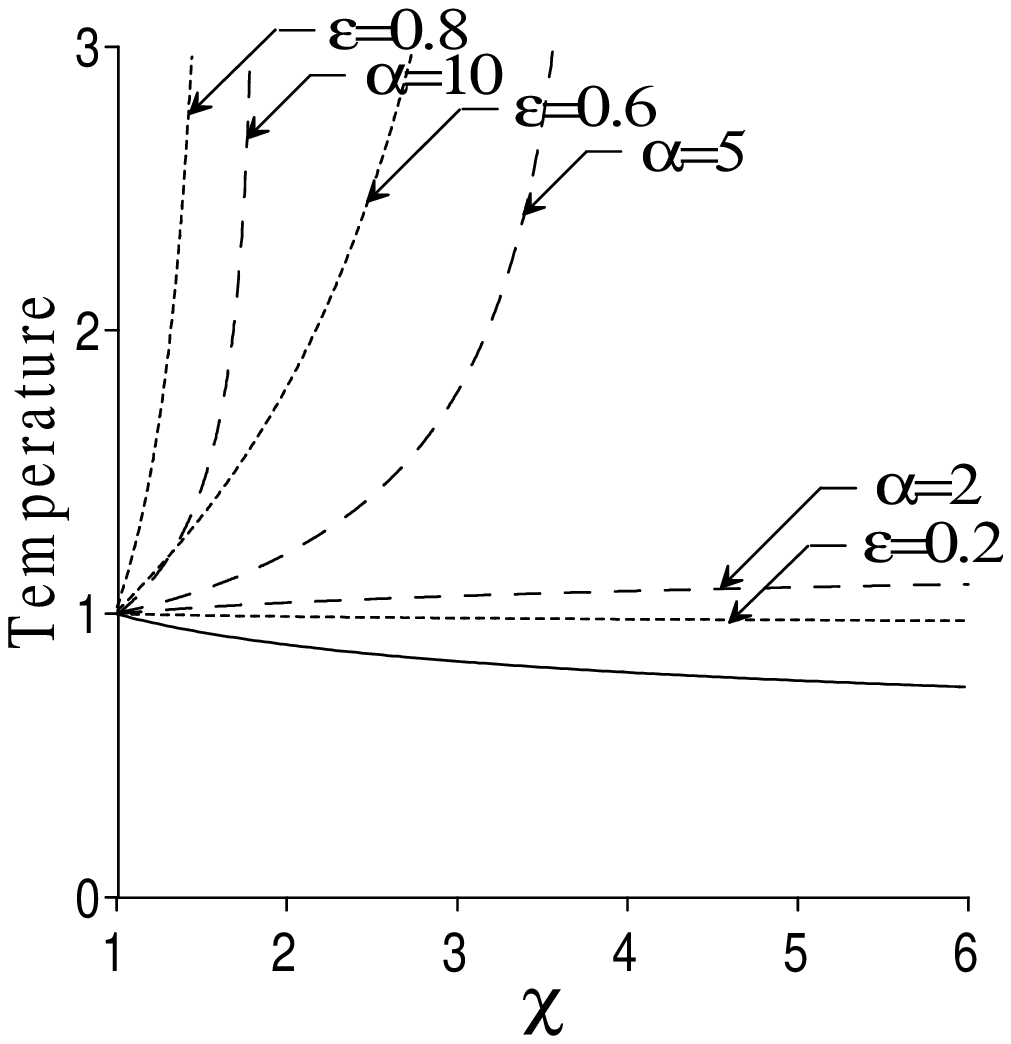}
} \caption{The distributions of the Lorentz factor, pressure and
particle density of the shocked gas as functions of self-similar
variable $\chi$. All quantities are normalized to unity at the
shock front where $\chi=1$. Long-dashed lines correspond to the
losses uniformly distributed in the downstream medium and
decreasing with time as $t^{-1}$. Short-dashed lines are for the
case of localized losses with $\delta {=}\: \eta {=}\: 0$. The
solid line --- solution of Blandford and McKee (1976).}
\end{figure}

In the case of localized losses we treat them as discontinuities
of the energy, momentum and particle number fluxes at the shock
front, which are characterized by three parameters $\varepsilon$,
$\delta$, $\eta$ equal to the fractions of corresponding fluxes
lost at the shock front in the front comoving frame. We obtain the
following expressions for the pressure $p_2$, number density $n_2$
and the Lorentz factor $\gamma_2$ immediately behind the shock
front:\\[-3\smallskipamount]
$$ p_2=\frac{2}{3}\xi_1\Gamma^2\:w_1,
\ \  \gamma_2^2=\frac{1}{2} \xi_2\Gamma^2, \ \
n_2=2\sqrt{2}\xi_3\Gamma\:n_1, \ \ \ \mbox{where}
\
\xi_1=\frac{3}{2} \frac{(1-\delta)\xi_2}{\xi_2^2-\xi_2+1},
$$\\[-1.8\bigskipamount]
$$%
\xi_2=1 + \frac{3(1-(1-\varepsilon)/(1-\delta))}
{1+\sqrt{4-3(1-\varepsilon)^2/(1-\delta)^2}}, \quad
\xi_3=\frac{(1-\eta)}{\sqrt{2}} \frac{\sqrt{\xi_2}}{1-\xi_2}.
$$%
They differ from the Taub adiabat only by numerical factors
$\xi_i$, which become unity in the absence of losses. Here $w_1 =
n_1 m_p c^2$ is the enthalpy density and $n_1$ the number density
of the external gas.

To find a self-similar solution we assume $\Gamma^2 {=}\:t^{-m}$
and introduce the similarity variable $\chi {=} \frac{ct-r}{ct-R}$,
where $r$ is distance from the center and $R$ the current radius
of the shock front. We find self-similar solutions in the case of
localized losses and in the case of $\varphi {=}
\frac{\alpha}{ct}$, $\alpha{=}const$, but they also exist for
non-uniform distributions of losses if $\alpha{=}\alpha(\chi)$.
The velocity, pressure and particle density are found in terms of
variables $\Gamma, \chi$.

From the energy balance equation we find that the power law index
$m$ is in the range $3 {\le} m {\le} 6$. There are two regions in
the parameter space where the solutions are qualitatively
different. In the case of small losses, $\alpha {<} 3$ or
$\varepsilon {+} \frac{2 - \sqrt{109}}{14} \delta {<} 1 {+}
\frac{2 - \sqrt{109}}{14}$, the index $m$ rises from 3 to 6 as the
losses increase. The pressure, Lorentz factor and number density
of the downstream are proportional to powers of the similarity
variable whose indices are different from those in the solution of
Blandford and McKee (1976). On the contrary, large losses lead to
the universal deceleration law of the shock: $m$ is equal to 6.
The problem is fully integrable but solutions can not be written
as explicit. In the high-loss solutions there appears an expanding
spherical cavity bounded by the contact discontinuity and the
temperature at its edge tends to infinity.

The solutions obtained are presented in Fig.1.

\section{Conclusion}

We have analyzed the dynamics of relativistic shock wave with
losses due to escape of neutral particles from plasma flow. Both
for localized and for distributed losses there are self-similar
solutions, which are different from those found previously for
lossless case. We find that increasing of the losses change the
dynamics of the shock deceleration qualitatively. In the case of
small losses, the role of ejected material asymptotically vanishes
and the Lorentz factor of the shock decreases as $t^{-m}$ with $m$
varying from 1,5 (no losses) to 3. In the opposite case of large
losses, the shock decelerates in accordance with universal law
$\Gamma \sim t^{-3}$ and the energy content in the ejecta
constitutes a significant fraction of the total energy budget.
Also, in the presence of large losses, the temperature and the
Lorentz factor of the fluid behind the shock can be non-monotonic
functions of distance from the shock.

\acknowledgments

This work was supported by the RFBR grants
nos. 02-02-16236 and 04-02-16987, the President of the Russian
Federation Program
for Support of Leading Scientific Schools (grant no.
NSh-1744.2003.2), and the program "Nonstationary Phenomena in
Astronomy" of the Presidium of the Russian Academy of Science.
E.V. Derishev acknowledges the support from the Russian Science
Support Foundation.


\begin{chapthebibliography}{1}

\bibitem[Baring \& Harding, 1995]{Baring}
Baring, M. G.; Harding, A. K., 1995, Adv. Sp. Res. 15(5), p.
153-156

\bibitem[(Blandford \& McKee, 1976)]{BM}
Blandford R. D., and McKee C. F., 1976, Phys. Fluids {\bf 19},
1130.

\bibitem[Derishev et al., 2003]{DKK}
Derishev, E. V.; Aharonian, F. A.; Kocharovsky, V. V.;
Kocharovsky, Vl. V., 2003, Phys. Rev. D 68, 043003

\bibitem[Meszaros, 2002]{Mesz}
Meszaros P., 2002, Ann. Rev. Astron. Astrophys. 40, p. 137-169

\bibitem[Piran, 2004]{Piran}
Piran, T., 2004, Rev. Mod. Phys., in press, astro-ph/0405503

\bibitem[(Taub, 1948)]{Taub}
Taub, A. H., 1948, Phys. Rev. 74, p. 328

\end{chapthebibliography}
\end{document}